\documentclass[12pt]{iopart}

\usepackage{iopams} 
\usepackage{graphicx}
\usepackage{hyperref}
\usepackage{graphicx}
\usepackage{subfigure}
\usepackage{arydshln}
\usepackage{inputenc}
\usepackage{graphicx}
\usepackage{hyperref}
\usepackage{cite}

\begin{document}
	
\title{On the evolution of the volume in Loop Quantum Cosmology}

\author{Beatriz Elizaga Navascu\'es}
\address{Department of Physics and Astronomy, Louisiana State University, Baton Rouge, LA 70803, U.S.A.}
\ead{bnavascues@lsu.edu}
\vspace{10pt}

\begin{abstract}
 
The dynamics of the expectation value of the volume is one of the key ingredients behind the replacement of the Big Bang singularity by a bounce in Loop Quantum Cosmology. As such, it is of great importance that this quantity is mathematically well-defined in the  space of physical states of the theory. A number of caveats have been raised about such a definition entering in conflict with the quantum evolution of states, motivated by the situation found in quantum geometrodynamics. We show that there are ways around these caveats, all of which are related to the existence of quantization prescriptions leading to a nondegenerate curvature operator in Loop Quantum Cosmology. Interestingly, the properties of the curvature operator that may allow for a good behavior of the volume are only possible thanks to the discreteness of the geometry characteristic of the loop quantization procedure.
 
\end{abstract}

\maketitle

\section{Introduction}\label{sec:Intro}

After decades of theoretical research, Loop Quantum Cosmology (LQC) \cite{abl,LQC2,ashparam} has become a mature field capable of addressing some of the fundamental questions about our Universe that lie beyond the applicability of General Relativity (GR). In particular, this discipline applies techniques from the canonical, nonperturbative, theory of Loop Quantum Gravity (LQG) \cite{LQG1,LQG2} to the quantization of cosmological spacetimes with a high degree of symmetry. With an understanding of the main features that this type of quantization brings, considerable efforts have been put into studying the effects that loop quantum gravitational phenomena could have had on the the primordial fluctuations in the Early Universe (see e.g. \cite{Bojo1,AAN,hyb-pert1,LQC3,Hybridreview,Ivanreview} and references therein). In fact, the field has developed to a state in which it has been possible to quantify potential imprints of LQG on our observations of the Cosmic Microwave Background (see e.g. \cite{hybpred1,AshtPRLLast,Ivanlast,hybothers,JCAPGB,JaviMerceRita}).

Perhaps the most remarkable prediction from LQC, which stands in contrast with results from traditional quantum geometrodynamics (also known as Wheeler--DeWitt formalism) \cite{WdW,QCHalliwell}, is the resolution of the Big Bang singularity via a bouncing mechanism of quantum origin \cite{APS1,APS2,MMO}. This bounce was originally obtained in homogeneous and isotropic cosmologies, via a combination of numerical techniques and theoretical knowledge of the Wheeler--DeWitt behavior of the LQC wavefunctions in the limit in which the scale factor is large. In this context, the bounce is often understood in terms of the variation of the volume of cubic cells with respect to a matter field (which plays the role of a relational time). The expectation value of this volume on some families of states follows the celebrated bouncing trajectories, that only differ from the relativistic ones when the energy density of the Universe approaches the Planck scale and can be well reproduced by effective equations \cite{APS1,Taveras,effparam,effash}. 

Despite its success, the lack of analytic derivations of the bounce directly from the quantum theory have motivated a number of caveats on its robustness \cite{madhavan,kaminski,critbojo,critlew}. A particularly worrisome one is the claim that the evolution of the volume operator as an expectation value may not be a well-defined quantity on the physical Hilbert space of the theory \cite{kaminski}, as this issue is actually found in quantum geometrodynamics \cite{madhavan,kaminski}. Were this claim turn out to be generically true for standard formulations of LQC, it could seriously affect the theoretical validity of the predictions extracted so far from the theory on realistic scenarios such as the physics of the Early Universe \cite{critlew}, where the quantum gravity effects on perturbations are codified via expectation values of geometric operators that often include the volume \cite{hybrMS,fluctdress}. The aim of this work is thus to study in detail if there is a genuine obstruction to the evolution of the volume in LQC, considering the ambiguities involved in the quantization procedure.

This article is organized as follows. In section \ref{sec:LQC} we summarize the loop quantization of homogeneous and isotropic spacetimes, and the role of dynamics in it. Section \ref{sec:Kaminski} reviews the line of reasoning leading to an obstruction in the definition of the evolution of the volume \cite{kaminski}. We provide ways of avoiding this obstruction by means of appropriate quantization choices in section \ref{sec:MMO}, and  we conclude our results and provide some outlook in section \ref{sec:conclusion}. We work in Planck units, with $c=\hbar=G=1$.

\section{Homogeneous and isotropic LQC}\label{sec:LQC}

The paradigmatic example for the application of loop techniques to the quantization of highly symmetric spacetimes is that of a homogeneous and isotropic cosmology minimally coupled to a massless scalar field $\phi$. In standard approaches to canonical quantum cosmology (including LQC), the only constraint that survives the symmetry reduction of GR is the Hamiltonian constraint \cite{abl}. Taking its density weight equal to one, the quantum representation of this constraint is given by an operator of the form

\begin{equation}
-\partial_{\phi}^2 - Q,
\label{eq:constr}
\end{equation}
defined on a kinematic Hilbert space, which we specify below in the case of LQC. In the equation above, $-i\partial_\phi$ is the momentum of the scalar field in the Schr\"odinger representation, and $Q$ is an operator that, basically, corresponds to the quantization of the (square of the) extrinsic curvature of the spatial hypersurfaces of the cosmological model. In both standard quantum geometrodynamics and LQC, $Q$ is (essentially) self-adjoint and positive, with an absolutely continuous spectrum equal to $[0,\infty)$ \cite{K&L_I}. 

Physical states of the quantum theory are those elements of (the algebraic dual of a dense subspace of) the kinematical Hilbert space that are annihilated by the constraint operator \cite{Dirac}. Typically, in LQC one restricts to solutions of the constraint equation which only contain positive frequencies of the scalar field $\phi$. Let us call $\mathcal{H}_{\mathrm{phys}}$ the physical Hilbert space of such solutions with positive frequency, endowed with an adequate choice of inner product \cite{LQC2}. Dynamics is defined there in a relational manner with respect to the scalar field $\phi$. Specifically, the evolution of physical states is realized as a map \cite{APS2}:

\begin{equation}
\psi \longmapsto \psi_{\phi}=e^{-i\phi\sqrt{Q}}\,\psi, \qquad \phi\in\mathbb{R},\qquad \psi\in\mathcal{H}_{\mathrm{phys}}.
\label{eq:pevol}
\end{equation}
Many of the physical predictions of LQC, such as the celebrated resolution of the Big Bang singularity via a bouncing mechanism and its consequences on the primordial inhomogeneities of our Universe, make use of the above notion of evolution to study the variation of physically relevant quantities. In particular, as we commented in the introduction, the bounce in LQC is broadly understood as arising from the dynamical behavior of the expectation value of the physical volume of cubic cells (or of the Universe itself, if it were spatially compact). The exact computation of the evolution of such an observable can only be addressed numerically, and the simulations show that the peaks of certain families of states follow bouncing trajectories that avoid the classical singularity \cite{APS1,APS2,effash}. In view of the numerical nature of these predictions, it is legitimate to ask the theory whether the expectation value of the volume operator $V_{\mathrm{LQC}}$ is, in fact, well-defined on the evolution of physical states, in the following sense. Let $D(O)$ denote the domain of any operator $O$ defined on $\mathcal{H}_{\mathrm{phys}}$. Then, given any $\psi\in D(V_{\mathrm{LQC}}^{1/2})$, does it also hold that $\psi_{\phi}\in D(V_{\mathrm{LQC}}^{1/2})$? The answer found e.g. in \cite{kaminski} is that, under certain assumptions, this only happens when $\psi=0$. If this result happened to be general in LQC, it would pose a serious drawback to the validity of the physical predictions extracted from this theory so far.

For clarity in the subsequent exposition, as well as to emphasize the importance of the volume as a fundamental operator in LQC, we end this section reviewing the explicit construction of the kinematic Hilbert space, together with the way in which $Q$ acts on it. In short, LQC is based on a noncontinuous quantization of the canonical pair formed by a variable $v$ and its momentum, where $|v|$ is proportional to the physical volume and its momentum encodes the information of the Ashtekar-Barbero connection. The kinematic Hilbert space of the theory is given by $\mathcal{H}_{\mathrm{kin}}\otimes L^2(\mathbb{R},d\phi)$, where \cite{abl,LQC2}

\begin{equation}
\mathcal{H}_{\mathrm{kin}}=\left\lbrace\psi:\mathbb{R}\rightarrow \mathbb{C}, \quad \sum_{v\in\mathbb{R}}|\psi(v)|^2 <\infty\right\rbrace
\end{equation}
is the so-called polymeric Hilbert space, on which the volume acts by multiplication: $V_{\mathrm{LQC}}\psi(v)\propto |v|\psi(v)$.  Due to the discreteness of the representation, in order to construct the operator $Q$ it is necessary to write its classical counterpart in terms of a circuit of holonomies of the Ashtekar-Barbero connection enclosing a minimum physical area, borrowing the techniques used in full LQG for similar means \cite{Thiemann_Reg}. The resulting operator has the following type of action on kinematic states for the geometry \footnote{In this work we focus on traditional approaches in LQC that regularize the curvature of the connection after fully performing a symmetry reduction on the Hamiltonian of the system.}:

\begin{equation}\label{eq:Qf}
-Q\psi(v)=f_{+}(v)\psi(v+4)-f_o(v)\psi(v)+f_{-}(v)\psi(v-4),
\end{equation}
where $f_{\pm}$ and $f_o$ are real functions whose explicit form we provide in the Appendix. Their properties depend on the specific prescription followed to construct $Q$. Indeed, the classical counterpart of this operator is a nonlinear function of the canonical variables employed in LQC, so its quantum representation is subject to (e.g. factor ordering) ambiguities. As we will see, important properties of the resulting physical Hilbert space, for instance in what concerns the domain of $V_{\mathrm{LQC}}^{1/2}$ under evolution, depend on how one fixes these ambiguities.

\section{The obstruction}\label{sec:Kaminski}

In this section we summarize the relevant arguments leading to the issue with $V_{\mathrm{LQC}}^{1/2}$ first noticed in \cite{kaminski} (see \cite{madhavan} for the first arguments pointing to the same result based on quantum geometrodynamics), when the operator $Q$ is defined using the so-called Ashtekar-Pawlowski-Singh (APS) prescription in LQC \cite{APS2,MOP,K&P}. This proposal was the first one leading to bouncing scenarios with acceptable physical properties. The reason why we are restricting here to the APS prescription is that the line of reasoning presented in \cite{kaminski} (leading to an obstruction to the evolution of the volume) does not necessarily apply when one considers other prescriptions for the construction of $Q$, as we will see.

In the APS model $Q=\Theta$, where $\Theta$ is an operator whose action leaves invariant any of the following separable subspaces of $\mathcal{H}_{\mathrm{kin}}$ (often called superselection sectors):

\begin{equation}\label{eq:superAPS}
\mathcal{H}_{\varepsilon}=\{\psi:\mathcal{L}_{\varepsilon}\rightarrow \mathbb{C}:\, \sum_{v\in \mathcal{L}_{\varepsilon}}|\psi (v)|^2 <\infty\}, 
\label{eq:HAPS}
\end{equation}
where $\mathcal{L}_{\varepsilon}=\{v=\varepsilon +4n,\,n\in\mathbb{Z},\, \varepsilon\in (0,4]\}$. Taking into account the large symmetry under the reflection $v\rightarrow -v$ that exists in the classical theory, as well as the spectral properties of $\Theta$, in the APS model $\mathcal{H}_{\mathrm{phys}}$ is identified with the Hilbert subspace of states $\psi$ in $\mathcal{H}_{\varepsilon}\oplus\mathcal{H}_{4-\varepsilon}$ such that $\psi(v)=\psi(-v)$, for any choice of superselection sectors with $\varepsilon \neq 2,4$ (we will comment on these exceptional cases later). The direct sum is needed to consider wavefunctions with support that is symmetric around $v=0$. Given any such choice of physical Hilbert space, the generalized eigenfunctions $e_k (v)$ of $\Theta$ satisfying $\Theta e_k(v)=k^2 e_k(v)$, with $k^2\in \mathbb{R}_{+}$ can be constructed in an iterative way via:

\begin{equation}\label{eq:iterativeAPS}
f_{+}(v)e_k (v+4)= -[k^2 -f_o (v)]e_k(v)-f_{-}(v)e_k (v-4) .
\end{equation}
In the APS construction, the functions $f_{\pm}$ are never zero when evaluated on any of the lattices $\mathcal{L}_{\varepsilon}$ or $\mathcal{L}_{4-\varepsilon}$ for $\varepsilon \neq 2,4$. Therefore, completely fixing each eigenfunction amounts to giving two pieces of initial data to the iterative relation above, e.g. the values $e_k(\varepsilon)$ and $e_k(\varepsilon - 4)$. It follows that the
generalized eigenspaces of $\Theta$ are two-fold degenerate: For each $k^2$, there exist two independent eigenfunctions $e_{|k|} (v)$ and $e_{-|k|}(v)$. In other words, functions of the form

\begin{equation}
\int_{\mathbb{R}} dk\, g(k) e_k (v),\qquad g\in C_0^{\infty}(\mathbb{R}\setminus \{0\}) 
\label{eq:eigenf}
\end{equation}
form a dense set in $\mathcal{H}_{\mathrm{phys}}$, where the integration over the whole of the real line makes the two-fold degeneracy of the operator $\Theta$ explicitly manifest. 

The fact that there are two eigenfunctions for any given (generalized) eigenvalue of the operator $\Theta$ has worrisome consequences in what respects the dynamical definition of the volume operator, as we see in the following.  First of all, one can show that $V_{\mathrm{LQC}}^{-1/2}e_k \in\mathcal{H}_{\mathrm{phys}}$ and that $e_k$ is analytic as a function of $k$ (see e.g. \cite{kaminski}), so the functions

\begin{equation}
h_{\psi}(k)=\langle V^{1/2}_{\mathrm{LQC}}\psi, V^{-1/2}_{\mathrm{LQC}} e_k\rangle,\qquad \psi\in D(V_{\mathrm{LQC}}^{1/2}),
\label{eq:fpsi}
\end{equation}
where $\langle .,.\rangle$ denotes the inner product in $\mathcal{H}_{\mathrm{phys}}$, are well-defined and analytic in $k$. Suppose that $\psi_\phi\in D(V_{\mathrm{LQC}}^{1/2})$. Then, the functions $h_{\psi_\phi}(k)$ must share these properties. Moreover, using the self-adjointness of $\sqrt{\Theta}$ and the dominated convergence theorem it is easy to see that

\begin{equation}
\int_{\mathbb{R}} dk\, g(k)h_{\psi_\phi}(k)=\int_{\mathbb{R}} dk\, g(k)e^{i\phi|k|}h_{\psi}(k),\qquad g\in C_0^{\infty}(\mathbb{R}\setminus \{0\}) .
\label{eq:intAPS}
\end{equation}
This implies simultaneously that $h_{\psi_\phi}(z)=e^{i\phi z}h_{\psi}(z)$ and $h_{\psi_\phi}(z)=e^{-i\phi z}h_{\psi}(z)$ almost everywhere in the complex plane, due to the analytic properties of the functions involved. Therefore, $h_{\psi}(z)=0$ almost everywhere. So, if $\psi\in D(V_{\mathrm{LQC}}^{1/2})$ and we assume that $\psi_\phi\in D(V_{\mathrm{LQC}}^{1/2})$, it follows necessarily that $\psi$ is orthogonal to the dense subspace of $\mathcal{H}_{\mathrm{phys}}$ spanned by functions of the form \eref{eq:eigenf}, namely, $\psi=0$. In other words, for any choice of $\mathcal{H}_{\mathrm{phys}}$ in the APS model with $\varepsilon\neq 2,4$, the expectation value of the volume operator fails to be well-defined on the evolution of physical states (other than the zero state).

It should be noted that exactly the same result is found if one follows the so-called simplified LQC (sLQC) prescription for the operator $Q$ \cite{sLQC,MOP,K&P}. This is because the degeneracy of this operator is again two-fold when its closure is defined on the sector of $\mathcal{H}_{\varepsilon}\oplus \mathcal{H}_{4-\varepsilon}$ that is symmetric under $v\rightarrow -v$, for any $\varepsilon \neq 2,4$ (see the Appendix).

\section{Avoidance of the issue}\label{sec:MMO}

There exist two particular identifications of the physical Hilbert space in the APS model for which the aforementioned issue with the evolution of the volume is avoided. They correspond precisely to the choices $\varepsilon=2,4$ for the superselection sectors \eref{eq:superAPS} that are invariant under the action of $\Theta$. These are quite exceptional inasmuch as $\mathcal{H}_{2}$ and $\mathcal{H}_{4}$ are the only sectors that contain wavefunctions that are symmetrically supported around $v=0$. This naturally leads to a direct identification of $\mathcal{H}_{\mathrm{phys}}$ with the Hilbert subspace of states $\psi$ in either $\mathcal{H}_{2}$ or $\mathcal{H}_{4}$ such that $\psi(v)=\psi(-v)$. In these respective cases, the action of $\Theta$ on any state involves evaluation of equation \eref{eq:Qf} on $v=\pm 2$ or $v=0,\pm 4$. But these points are especial precisely because the only zeros of $f_\pm (v)$ and $f_o(v)$ are respectively located at $v=0,\mp 2,\mp 4$ and $v=0$. A look at the iterative equation \eref{eq:iterativeAPS} then shows that generalized eigenfunctions are completely fixed by a single datum, $e_k(2)=e_k(-2)$ or $e_k(4)=e_k(-4)$, so the absolutely continuous spectrum of $\Theta$ is nondegenerate. In other words, we now have that functions obtained after replacing $\mathbb{R}$ by $\mathbb{R}_{+}$ in \eref{eq:eigenf} form a dense set in $\mathcal{H}_{\mathrm{phys}}$. One can then easily show that, in this case,

\begin{equation}
\int_{0}^{\infty} dk\, g(k)h_{\psi_\phi}(k)=\int_{0}^{\infty} dk\, g(k)e^{i\phi k}h_{\psi}(k),\qquad g\in C_0^{\infty}(\mathbb{R}\setminus \{0\})
\label{eq:intAPSgood}
\end{equation}
which implies only that $h_{\psi_\phi}(z)=e^{i\phi z}h_{\psi}(z)$ almost everywhere. This equality alone is insufficient to conclude that $h_{\psi}(z)=0$ almost everywhere. Therefore the obstruction found in the previous section (and originally pointed out in \cite{madhavan,kaminski}) to the evolution of the volume operator on physical states is directly avoided in the APS model if one restricts all attention to any of the exceptional superselection sectors $\varepsilon=2,4$.

In principle, one could put into question the robustness of the loop quantization procedure in cosmology if the validity of its predictions (at least in what concerns the volume) drastically depends on the choice of superselection sector. This is especially so since, from a physical perspective, there is no preferred choice a priori. However, one can actually consider an alternative prescription for the geometric operator $Q$ that does not suffer at all from the studied issue about the evolution of the volume on physical states. This model was originally proposed by the authors Mart\'in-Benito, Mena Marug\'an and Olmedo (MMO) as an isotropic limit of the loop quantization of Bianchi I cosmologies, and it takes into account that the sign of the triad operators in LQC does not commute with the basic holonomy operators that appear in the Hamiltonian constraint \cite{MMO}.

In the MMO model, we have $Q=\hat{\Omega}^2$ with an operator $\hat{\Omega}^2$ that simply differs from the APS one $\Theta$ in a compact perturbation, yet this difference results into some especially appealing properties. Specifically, the MMO operator $\hat\Omega^2$ is such that the functions $f_{+}$ and $f_{-}$ are zero when evaluated on, respectively, $[-4,0]$ and $[0,4]$. Furthermore, $f_o(0)=0$ (see the Appendix for details). It follows that its action separates the kinematical Hilbert space of LQC into the following superselection sectors:  

\begin{equation}\label{eq:superMMO}
\mathcal{H}^{\pm}_{\varepsilon}=\{\psi:\mathcal{L}^{\pm}_{\varepsilon}\rightarrow \mathbb{C}:\, \sum_{v\in\mathcal{L}^{\pm}_{\varepsilon}}|\psi (v)|^2 <\infty\}, 
\label{eq:HMMO}
\end{equation}
where $\mathcal{L}^{\pm}_{\varepsilon}=\{v=\pm(\varepsilon +4n),\,n\in\mathbb{N},\, \varepsilon\in (0,4]\}$. In particular, states supported on positive and negative values of $v$ never get related by the action of the operator and states with support in $v=0$ are completely decoupled from the theory. The physical Hilbert space $\mathcal{H}_{\mathrm{phys}}$ of the model can then be identified with any of these superselection sectors. On each one of them, $\hat\Omega^2$ is essentially self-adjoint with an absolutely continuous and nondegenerate spectrum equal to $[0,\infty)$. In other words, its generalized eigenfunctions satisfying $\hat{\Omega}^2 e_k(v)=k^2 e_k(v)$ provide a resolution of the identity in $\mathcal{H}_{\mathrm{phys}}$ when $k$ runs in (e.g.) $\mathbb{R}_{+}$. The nondegeneracy of the spectrum directly follows from adapting the iterative relation \eref{eq:iterativeAPS} to the present case, namely, using the functions $f_{\pm}$ and $f_o$ corresponding to $\hat{\Omega}^2$. Indeed, solutions are totally fixed given the single initial datum $e_k(\pm\varepsilon)$.  A completely analogous argument as the preceeding one for the APS especial cases $\varepsilon=2,4$ then shows that the obstruction found for the evolution of the volume on physical states is completely absent if one employs the MMO prescription to construct the operator $Q$ in LQC, regardless of the choice of superselection sector.

For completeness, we end this section noticing that the aforementioned issue is avoided as well in the especial cases $\varepsilon=2,4$ of the sLQC model, and in the so-called sMMO prescription \cite{MOP,MerceV} for any choice of its superselection sectors (which have the same form as in the MMO model). We refer again to the Appendix for details.

\section{Conclusions}\label{sec:conclusion}

In this work we have addressed the question of whether there exists a genuine obstruction to the evolution of the expectation value of the volume operator in LQC. In the last years this question has been raised as a possible caveat to standard formulations of the theory \cite{kaminski,critbojo,critlew}, partially motivated by the fact that there actually exist issues with the evolution of the scale factor in more traditional Wheeler--DeWitt models \cite{madhavan}. We have shown that, even though an obstruction may arise if one is not careful enough when constructing the quantum theory, it can be easily avoided by fixing some quantization ambiguities appropriately. In particular, there is no such issue whatsoever if one studies the MMO and sMMO models, and neither if one focuses on the especial superselection sectors $\varepsilon=2,4$ of the APS and sLQC models.

Interestingly, the existence or avoidance of the aforementioned obstruction is tightly linked to the degeneracy of the spectrum of the operator $Q$ that represents the extrinsic curvature of the cosmological model. In LQC this is always a finite difference operator of second order, due to the discreteness of the quantum geometry inherited from LQG. Such discreteness makes it possible that, for adequate quantization choices, the generalized eigenfunctions of $Q$ may be completely fixed in terms of just one datum (even if the finite difference equation that they must obey is of second order). This results into a nondegenerate spectrum for $Q$, which in turn is necessary for the evolution of the volume to be well-defined on physical states. In this context, it is worth mentioning that this situation cannot be found if one follows a traditional Wheeler--DeWitt quantization of the system. Indeed, in that case the continuity of the metric variables does not allow for a nondegenerate operator $Q$ which, furthermore, decouples the zero-volume state from the theory.

We should note that the avoidance of the obstruction under study via the nondegeneracy of the spectrum of $Q$ may not be, in principle, sufficient to guarantee that the evolution of the volume is well-behaved. An analytical study of this issue is  beyond the scope of this article at present, due to the lack of sufficient analytic knowledge on the large-volume behavior of states under evolution. In this context, it would be particularly insightful to perform a numerical analysis that does not adhere to the truncations in momentum space that have been traditionally used to derive the bounce in LQC \cite{APS2,MOP}.

Finally, it would be interesting to see if results similar to the ones found here carry over to more recent proposals for the quantum representation of the Hamiltonian constraint in LQC, that follow alternative regularization procedures in what respects the dependence on the Ashtekar-Barbero connection \cite{Ma,DaporLieg,DaporLieg2,Alex,PmLQC}. These models  have received  considerable attention lately in what respects the study of primordial fluctuations \cite{hybothers,AlexLQC,Ivanreview}. Moreover, the insights gained here about the role of the degeneracy of relevant geometric operators in the theory could prove useful in the study of interior black hole models, which have become an object of renewed research in the LQC community \cite{AOS,AOS2,Cong1,Cong2,AlexBH,AndresBH}.

\section*{Acknowledgments}
The author is very grateful to I. Agullo, A. Delhom, J. Lewandowski, G.A. Mena Marug\'an, J. Olmedo, T. Pawłowski, and M. Varadarajan for discussions. This work was supported by Grants NSF-PHY-1903799 and NSF-PHY-2206557, and by funds of the Hearne Institute for Theoretical Physics. It was also partially supported by Project No. MICINN PID2020-118159GB-C41 from Spain.

\appendix
\setcounter{section}{1}

\section*{Appendix}

In this Appendix we provide the explicit definition of the functions $f_{\pm}$ and $f_o$, that determine the action of the operator $Q$ in LQC via equation \eref{eq:Qf}. We recall that the form and properties of these functions depend on which prescription is chosen for the construction of $Q$. In this work, we distinguish between the following cases:
\begin{itemize}
\item[i)] The APS model is characterized by the functions \cite{APS2,MOP,K&P}:

\begin{equation}\fl
f_{\pm}(v)=\mathfrak{B}(v\pm 4)^{1/2}\tilde{f}(v\pm 2)\mathfrak{B}(v)^{1/2},\quad f_o(v)=\mathfrak{B}(v)\left[f_{+}(v)+f_{-}(v)\right]
\end{equation}
where we have defined

\begin{equation}\fl
\mathfrak{B}(v)=\frac{27}{8}|v|\left||v+1|^{1/3}-|v-1|^{1/3}\right|^{-3},\quad \tilde{f}(v)=\frac{3\pi}{8}|v|\left||v+1|-|v-1|\right|,
\end{equation}
for any $v\neq 0$, and $\mathfrak{B}(0)=\tilde{f}(0)=0$ \footnote{In this work we choose to define the value of the function $\mathfrak{B}(v)$ as in \cite{MOP} to explicitly avoid any problems at $v=0$ (see e.g. \cite{KLP} for other ways around this issue in the APS and sLQC models).}. From these expressions it is clear that the zeros of $f_{\pm}(v)$ occur at $v=0,\mp 2,\mp 4$, whereas $f_o (v)$ only vanishes at $v=0$.

\item[ii)] In sLQC we have the following, simpler, functions \cite{MOP,K&P}:

\begin{equation}
f_{\pm}(v)=\frac{3\pi}{4}\sqrt{v|v\pm 4|}|v\pm 2|,\quad f_o(v)=\frac{3\pi }{2}v^2,
\end{equation}
which vanish exactly at the same respective values of $v$ as the APS ones.
\item[iii)] The functions of the MMO model are given by \cite{MMO}
\begin{eqnarray}
f_{\pm}(v)=\frac{\pi}{12}\tilde{g}(v\pm 4)s_{\pm}(v\pm 2)\tilde{g}(v\pm 2)^2 s_{\pm}(v)\tilde{g}(v) \nonumber \\ f_o(v)=\frac{\pi}{12}\tilde{g}(v)^2\left[\tilde{g}(v+2)^2s_{+}(v)^2+\tilde{g}(v-2)^2s_{-}(v)^2\right],
\end{eqnarray}
where we have defined

\begin{equation}\fl
\tilde{g}(v)=\left||1+v^{-1}|^{1/3}-|1-v^{-1}|^{1/3}\right|^{-1/2},\quad s_{\pm}(v)=\mathrm{sign}(v\pm 2)+\mathrm{sign}(v)
\end{equation}
for any $v\neq 0$, and $\tilde{g}(0)=0$. Clearly, we have that $f_{+} (v)$ and $f_{-}(v)$ vanish in the respective intervals $[-4,0]$ and $[0,4]$, whereas $f_o(v)$ is only zero at $v=0$.

\item[(iv)] Finally, sMMO is characterized by the simpler functions \cite{MOP}:
\begin{eqnarray}
f_{\pm}(v)=\frac{3\pi}{16}\sqrt{v|v\pm 4|}|v\pm 2|s_\pm(v\pm2)s_\pm (v),\nonumber \\ f_o(v)=\frac{3\pi }{16}|v|\left[|v+2|s_{+}(v)^2+|v-2|s_{-}(v)^2\right],
\end{eqnarray}
which vanish at the same values and intervals of $v$ as the MMO ones.

\end{itemize}

The superselection sectors of the theory, that are preserved by the action of $Q$ in each case, follow directly from the indicated properties of the functions $f_{\pm}$ and $f_o$. The same happens with the multiplicity of the spectrum of this operator in each sector.

\end{document}